# The effects of high-velocity supernova kicks on the orbital properties and the sky distributions of neutron-star binaries

**Niel Brandt and Philipp Podsiadlowski**

*Institute of Astronomy, Madingley Road, Cambridge CB3 0HA*

**ABSTRACT**

We systematically investigate the effects of high supernova kick velocities on the orbital parameters of post-supernova neutron-star binaries. Using Monte-Carlo simulations, we determine the post-supernova distributions of orbital parameters (orbital period, eccentricity, system velocity, spin inclination, ratio of spin to orbital angular momentum) for progeneitors of high-mass X-ray binaries (HMXBs) and low-mass X-ray binaries (LMXBs). With the recent distribution of pulsar birth velocities by Lyne & Lorimer (1994), only about 27 % of massive systems remain bound after the supernova, of which $\sim 26\,\%$ immediately experience dynamical mass transfer and possibly merge to become Thorne-Żytkow objects. The correlations between various orbital parameters can be compared with observational samples to yield information about supernova kick velocities and pre-supernova orbital-period distributions. After the supernova, the spins of most stars in massive systems have large inclinations with respect to their orbital axes, and a significant fraction of systems ($\sim 20\,\%$) contain stars with retrograde spins. This may have important implications for the interpretation of those HMXBs which seem to have tilted, 'precessing' accretion disks. We estimate that the spin angular momentum in the massive components of most HMXBs is a significant fraction $(0.1 - 0.4)$ of the total orbital angular momentum. Therefore spin – orbit coupling effects may be important in many HMXBs.

In the case of low-mass companions, we find that $\sim 19\,\%$ of systems remain bound after the supernova, of which $\sim 57\,\%$ experience immediate dynamical mass transfer. The systems that survive as binaries and become LMXB progenitors attain a very large system velocity of $180 \pm 80\,\mathrm{km\,s^{-1}}$ after the supernova. There is a relatively tight correlation between the eccentricity and the post-supernova orbital period in these systems. All LMXBs with post-supernova periods longer than a few days initially have very large eccentricities. This may suggest that there should be a special class of LMXBs with periodic outbursts, of which Cir X-1 may be an extreme representative. We also use the results of these calculations to simulate the sky distributions of HMXBs and LMXBs. The simulated distributions agree with observed samples. Most importantly, the distribution of Galactic LMXBs is consistent with an ordinary Galactic disk population which has been widened because of large supernova kicks and does not require a special population of progenitors. The observed LMXB distribution does not provide a strong constraint on the age of LMXBs since the supernova, although there may be weak hint that they are relatively young, with ages less than $\sim 10^8\,\mathrm{yr}$.

# 1 INTRODUCTION

It has long been known that a small asymmetry during a supernova explosion can impart a substantial kick to a newborn neutron star (e.g. Shklovskii 1969; Sutantyo 1978; Dewey & Cordes 1987; Bailes 1989). For example, a one per cent asymmetry in the neutrino momentum flux is sufficient to give a neutron star a kick velocity of $\sim 400\,\mathrm{km\,s^{-1}}$. However, while a variety of kick mechanisms have been proposed in the past, the detailed physics of the kick process has remained very poorly understood (e.g. Harrison & Tademaru 1975; Woosley 1987; Burrows & Fryxell 1992; Duncan & Thompson 1992; Herant, Benz & Colgate 1992; Janka & Müller 1994). Recent measurements of pulsar proper motions (Bailes et al. 1989; Fomalont et al. 1992; Harrison, Lyne & Anderson 1993), the realization that some previous pulsar velocities were systematically low (Harrison & Lyne 1993) and the adoption of a new distance scale for pulsars (Taylor & Cordes 1993) have led Lyne & Lorimer (1994) to reassess pulsar space velocities. From a sample of young pulsars they derive a mean pulsar birth velocity of $450 \pm 90\,\mathrm{km\,s^{-1}}$, substantially higher than previous estimates. Analyses of possible associations between pulsars and supernova remnants also suggest stronger kicks, with mean pulsar velocities as high as $500\,\mathrm{km\,s^{-1}}$ or higher (Caraveo 1993; Harrison, Lyne & Anderson 1993; Frail, Goss & Whiteoak 1994). This is also the case for the soft gamma-ray repeaters SGR $0525 - 66$ and SGR $1806 - 20$ which have high inferred space velocities ($> 1200$ km s$^{-1}$ and $> 500$ km s$^{-1}$, respectively), possibly implying unusual manners of birth (Kulkarni et al. 1994; Rothschild, Kulkarni & Lingenfelter 1994). The bizarre and violent X-ray binary Cir X-1 also appears to have a high space velocity. Its system velocity is larger than $\sim 200\,\mathrm{km\,s^{-1}}$ (Duncan, Stewart & Haynes 1993) and may be larger than $\sim 450\,\mathrm{km\,s^{-1}}$ if one accepts its speculated association with the supernova remnant $G321.9 - 0.3$ (Haynes et al. 1978; Haynes et al. 1986; Stewart et al. 1993).

An increase of this size in natal kick strength will have a major effect on our understanding of the fraction of pulsars that remain bound in binary systems, globular clusters and the Galaxy. In addition, stronger supernova kicks lead to significant changes in the orbital properties of low-mass X-ray binaries (LMXBs) and high-mass X-ray binaries (HMXBs). The effect of kicks on X-ray binaries has been studied previously for kick velocities of order $100-200$ km s$^{-1}$ (Flannery & van den Heuvel 1975; De Cuyper et al. 1976; Amnuel & Guseinov 1976; Sutantyo 1978; Hills 1983; Sutantyo, van der Linden & van den Heuvel 1986; Wijers, van Paradijs & van den Heuvel 1992). As kick velocities increase, fewer systems remain bound (Hills 1983; van den Heuvel 1994). Systems that suffer stronger natal kicks have generally higher space velocities and are more likely to have misaligned spin axes relative to their orbital axes. The fraction of systems with retrograde spins relative to the orbital motion and the percentage of bound systems that could lead to the formation of Thorne-Żytkow objects (TŻOs) increase with increasing kick velocity. In this paper we investigate in a systematic fashion the effects of high supernova kick velocities on the properties of post-supernova binaries. In Section 2 we summarize important relations between the pre-supernova and post-supernova orbital parameters and illustrate these for a typical HMXB progenitor. In section 3 we consider the distribution of post-supernova orbital parameters for HMXB and LMXB progeneitors with increasing sophistication. In Section 4 we examine the implications of high kick velocities on the sky distributions of HMXBs and LMXBs, and in Section 5 we discuss our results.

# 2 THE EFFECTS OF ASYMMETRIC SUPERNOVAE ON BINARY SYSTEMS

## 2.1 Post-supernova relations

The post-supernova parameters of a binary in which one star experiences an asymmetric supernova explosion have been calculated by many authors in the past (e.g. Sutantyo 1978; Dewey & Cordes 1987; Verbunt et al. 1990, Wijers et al. 1992). We will therefore only present the main relations between initial and final binary parameters without detailed derivations.

We consider a pre-supernova binary consisting of two stars (1 and 2) with masses $M_1$ and $M_2$, respectively, in a circular orbit with orbital separation $a$ and orbital period $P$. The initial velocities of the two stars in the center-of-mass (CM) frame are $\mathbf{v_1}$ and $\mathbf{v_2}$, respectively. Star 1 experiences a supernova which (instantaneously) produces a remnant of mass $M_1'$. In addition, the remnant receives a kick with a velocity $\mathbf{v}_{\text{kick}}$ due to an asymmetry in the explosion. We do not consider the effects of the supernova shell on the companion star (Colgate 1970; Wheeler, Lecar & McKee 1975), since these will generally be small (Fryxell & Arnett 1981). The direction of the kick is specified by two angles $\theta$ and $\phi$. The angle $\theta$ specifies the angle between the direction of the kick-velocity vector and the initial orbital plane, and $\phi$ is the angle between the initial direction of motion of star 1 and the projection of the kick-velocity vector onto the orbital plane. The overall geometry of the problem is shown in Fig. 1. We find it convenient to use non-dimensional masses and velocities by defining

$$\tilde{m} = \frac{M_1 + M_2}{M_1' + M_2}, \tag{2.1}$$

and

$$\tilde{v} = \frac{v_{\text{kick}}}{v_{\text{orb}}}, \tag{2.2}$$

where $v_{\text{kick}} = |\mathbf{v_{kick}}|$ and $v_{\text{orb}}$ is the initial relative orbital velocity:

$$v_{\text{orb}}^2 = \frac{G(M_1 + M_2)}{a} \tag{2.3a}$$

or by Kepler's law

$$v_{\text{orb}}^3 = \frac{2\pi G(M_1 + M_2)}{P} \tag{2.3b}$$

($G$ is the gravitational constant).

With these definitions, we can write the post-supernova energy (in the new CM frame) as

$$E' = -\frac{GM_1'M_2}{2a}\left\{2 - \tilde{m}\left[1 + 2\tilde{v}\cos\phi\cos\theta + \tilde{v}^2\right]\right\}. \tag{2.4}$$

Since $E'$ can also be written as $E' = -GM_1'M_2/2a'$, where $a'$ is the post-supernova semi-major axis of the binary, equation (2.4), in effect, gives $a'$ and with Kepler's law the new orbital period $P'$. For the system to remain bound after the supernova, the energy $E'$ has to be negative. Inspection of equation (2.4) shows that this requires that $\tilde{v}$ is less then $\tilde{v}_{\max}$, where

$$\tilde{v}_{\max} = 1 + \sqrt{\frac{2}{\tilde{m}}}. \tag{2.5}$$

In addition, if $\tilde{m} > 2$, $\tilde{v}$ has to be larger than $\tilde{v}_{\min}$, where

$$\tilde{v}_{\min} = 1 - \sqrt{\frac{2}{\tilde{m}}}. \tag{2.6}$$

The general requirement on the kick directions for a bound post-supernova orbit is

$$\cos\phi\cos\theta < \frac{1}{2\tilde{v}}\left[\frac{2}{\tilde{m}} - 1 - \tilde{v}^2\right]. \tag{2.7}$$

The expression for the post-supernova eccentricity $e$ can be written as

$$e^2 = 1 - \tilde{m}\left\{2 - \tilde{m}\left[1 + 2\tilde{v}\cos\phi\cos\theta + \tilde{v}^2\right]\right\}\left[(1 + \tilde{v}\cos\phi\cos\theta)^2 + (\tilde{v}\sin\theta)^2\right]. \tag{2.8}$$

One can show that equation (2.8) implies that there is a maximum $\tilde{v}$ for which the post-supernova binary can be circular, which is given by

$$\tilde{v}_{\max,\text{circ}} = 1 + \sqrt{\frac{1}{\tilde{m}}}. \tag{2.9}$$

We will also need an expression for the system velocity, $v_{\text{sys}}$, the binary system receives as a result of the explosion. This is just the velocity of the post-supernova CM relative to the initial CM frame:

$$v_{\text{sys}} = \frac{v_{\text{orb}}}{M_1' + M_2}\left\{\left(\frac{\mu\Delta M_1}{M_1}\right)^2 - 2\frac{\mu\Delta M_1 M_1'}{M_1}\tilde{v}\cos\phi\cos\theta + (M_1'\tilde{v})^2\right\}^{1/2}, \tag{2.10}$$

where $\mu \equiv M_1M_2/(M_1 + M_2)$ and $\Delta M_1 \equiv M_1 - M_1'$. Equation (2.10) implies that the range of $v_{\text{sys}}$ is bounded by

$$\frac{\left|\frac{\mu\Delta M_1}{M_1} - M_1'\tilde{v}\right|}{M_1' + M_2} \leq \frac{v_{\text{sys}}}{v_{\text{orb}}} \leq \frac{\frac{\mu\Delta M_1}{M_1} + M_1'\tilde{v}}{M_1' + M_2}, \tag{2.11}$$

although a tighter lower bound can be obtained via examination of equation (2.10) keeping equation (2.7) in mind. An important consequence of equation (2.11) is that there is a maximum system velocity a post-supernova binary can have, since $v_{\mathrm{orb}}$ is constrained by the requirement that in the pre-supernova system neither component overfills its Roche lobe. The condition that star 2 fills its Roche lobe exactly implies a maximum initial orbital velocity

$$v_{\mathrm{orb,max}} = \sqrt{\frac{G(M_1 + M_2)}{R_2} f\left(\frac{M_1}{M_2}\right)}, \qquad (2.12)$$

where $R_2$ is the radius of star 2, and $f$ the ratio of star 2's Roche-lobe radius to the orbital separation given by (Eggleton 1983)

$$f(q) = \frac{0.49\, q^{-2/3}}{0.6\, q^{-2/3} + \ln\left(1 + q^{-1/3}\right)}. \qquad (2.13)$$

Thus, for a given $v_{\mathrm{kick}}$, the maximum post-supernova system velocity becomes

$$v_{\mathrm{sys,max}} = \frac{\frac{\mu \Delta M_1}{M_1} v_{\mathrm{orb,max}} + M_1' v_{\mathrm{kick}}}{M_1' + M_2}. \qquad (2.14)$$

Since there is also a maximum kick velocity for which the orbit remains bound (eq. 2.5), the maximum post-supernova system velocity for arbitrary kick velocities can be written as

$$v_{\mathrm{sys,max}} = \frac{\frac{\mu \Delta M_1}{M_1} + M_1'\left(1 + \sqrt{\frac{2}{\tilde{m}}}\right)}{M_1' + M_2}\, v_{\mathrm{orb,max}}. \qquad (2.15)$$

Equation (2.15) can be used to constrain the properties of pre-supernova systems from measured post-supernova system velocities.

Finally, the angle $\nu$ of the post-supernova orbital-angular-momentum vector with respect to the initial one can be written as

$$\cos \nu = \frac{1 + \tilde{v} \cos \phi \cos \theta}{\left[\tilde{v}^2 \sin^2 \theta + (1 + \tilde{v} \cos \phi \cos \theta)^2\right]^{1/2}}. \qquad (2.16)$$

Since the systems we consider have experienced at least one phase of mass transfer before the supernova, the spin vectors of the two stars are likely to be aligned with the orbital-angular-momentum vector before the supernova. We assume that the spin directions are not affected by the supernova. This is certainly a good assumption for the companion star, since it will be little affected by the supernova (Fryxell & Arnett 1981), but need not be correct for the newborn neutron star, in particular since it is not clear whether the core of the exploding star will have been aligned with the massive envelope before the supernova. This depends on the poorly understood coupling between the core and the envelope of the neutron-star progenitor (S. Thorsett, private communication). With these assumptions, $\nu$ gives the inclination of the stars' spin vectors with respect to the orbital-angular-momentum vector after the supernova. The spins are retrograde with respect to the orbital motion if $\nu > 90°$. Equation (2.16) shows that this will be the case if

$$\cos\phi\cos\theta < -\frac{1}{\tilde{v}}. \tag{2.17}$$

Therefore, a necessary but not sufficient requirement for a retrograde post-supernova spin of the two stars is that $\tilde{v} > 1$ or (obviously) $v_{\text{kick}} > v_{\text{orb}}$.

## 2.2 Probabilities of bound and retrograde orbits

The probability that a binary remains bound or attains retrograde spins depends on $v_{\text{kick}}$ and the distribution of kick angles $\phi$ and $\theta$. In the case that all kick directions are equally probable, the probability that

$$\cos\phi\cos\theta < A, \quad -1 \leq A \leq +1 \tag{2.18}$$

(see eqs 2.7 and 2.17), is given by

$$P(\cos\phi\cos\theta < A) = \frac{1}{2}(1 + A). \tag{2.19}$$

In Fig. 2 we plot the probability for bound orbits (combining eqs 2.7 and 2.19) as a function of $v_{\text{kick}}/v_{\text{orb}}$ for various values of $\tilde{m}$. The fraction of bound post-supernova binaries with retrograde spins becomes, for $\tilde{v} > 1$,

$$P(\text{bound, retrograde}) = \frac{1 - \frac{1}{\tilde{v}}}{1 + \frac{1}{2\tilde{v}}\left[\frac{2}{\tilde{m}} - 1 - \tilde{v}^2\right]}. \tag{2.20}$$

This fraction reaches 100 % at

$$\tilde{v}_{\text{retro},100\%} = \sqrt{1 + \frac{2}{\tilde{m}}}. \tag{2.21}$$

The value of $\tilde{v}$ at which 50 % of post-supernova binaries have retrograde spins is

$$\tilde{v}_{\text{retro},50\%} = \sqrt{4 + \frac{2}{\tilde{m}}} - 1. \tag{2.22}$$

In Fig. 3a we indicate the various regimes for bound orbits and orbits with retrograde/prograde spins in a $\tilde{v} - \tilde{m}$ plane. In Fig. 3b we show the same regimes in a $v_\text{kick} - P_\text{orb}$ plane for a total initial mass of 20 $M_\odot$ and a total post-supernova mass of 16.4 $M_\odot$ (see Section 2.3). Another useful reference for kick induced retrograde orbits is De Cuyper (1984).

If the distribution of kick directions is not isotropic, these probabilities will generally be quite complicated. In some theoretical models for the supernova kick, one might expect the kick to be preferentially along the initial spin or magnetic-field axis (e.g. Harrison & Tademaru 1975; Duncan & Thompson 1992), although this is presently not supported by observations (Anderson & Lyne 1983). To examine this possibility, we also calculated the probability for $P(\cos\phi\cos\theta < A)$ if the range of kick directions is restricted to a cone with cone angle $\alpha$ with respect to the initial spin axis, where all directions within the cone are equally probable. We find

$$P(\cos\phi\cos\theta < A) = \frac{1}{[\pi(1-\cos\alpha)]} \left\{ \frac{\pi}{2}(1-\cos\alpha + A) + A\arctan\left[\frac{\cos\alpha}{B}\frac{A}{|A|}\right] \right.$$
$$\left. -\cos\alpha \arctan\frac{A}{B} - \arctan\frac{|A|\cos\alpha}{B} \right\}, \qquad (2.23)$$

where
$$B = \sqrt{\sin^2\alpha - A^2}.$$

Note that, in this case, retrograde post-supernova spins are only possible if

$$\sin\alpha > \frac{1}{\tilde{v}}. \qquad (2.24)$$

### 2.3 Application to HMXB progeneitors

To illustrate these relations we consider a set of typical initial parameters appropriate for the progenitor of a HMXB before the first supernova. We consider an initially circular binary system in which star 1 has an initial mass of 5 $M_\odot$ and star 2 a mass of 15 $M_\odot$. These parameters are typical for a system which has experienced one phase of mass transfer (conservative or dynamical) in which the initially more massive star (star 1) lost or transferred a large fraction of its hydrogen-rich envelope and has become a helium star (for a review see Bhattacharya & van den Heuvel 1991). Star 1 explodes in an asymmetric supernova leaving behind a 1.4 $M_\odot$ neutron star which has received a kick of 450 km s$^{-1}$. $\tilde{m} = 1.22$ and $\mu = 3.75\,M_\odot$.

In Fig. 4 we show the distributions of final orbital period, final eccentricity and CM velocity as a function of initial period. A natural lower boundary to the initial period exists since the companion star's radius cannot exceed its effective Roche lobe radius (eq. 2.13), and we draw this boundary as a solid line (the stellar radius is taken from Podsiadlowski [1989]). In addition, equation (2.5) (combined with eq. 2.3b) implies that, for given masses and kick velocity, there are no bound orbits above a certain maximum initial period, in this case 25.13 days. Below an initial period of 2.12 days, all orbits must be prograde while above 9.09 days all orbits must be retrograde (see Sec. 2.2). For initial periods of less than about 10 days, Fig. 4a shows that the median final period is reasonably close to the initial period. There is a highly asymmetric spread about the median final period values. Fig. 4b illustrates the wide range of possible final eccentricities below about 15 days and the significantly narrower range above this initial period. The sharp rise in the eccentricity distribution occurs at the initial period above which there are no low-eccentricity post-supernova systems (as given by eq. 2.9). Fig. 4c illustrates the relatively narrow range of allowed post-supernova system velocities, where the minimum system velocity (which is 70.8 km s$^{-1}$ for an initial period of 25.13 days) is given by equations (2.7), (2.10) and (2.11) and the maximum by equations (2.12) – (2.14).

# 3 THE DISTRIBUTION OF POST-SUPERNOVA ORBITAL PARAMETERS

We now turn to an examination of the distributions of orbital properties after the supernova kick. In Section 3.1 we start with two simple examples to illustrate the dependence of the post-supernova parameters on $v_{\text{kick}}$. In Section 3.2 we use more realistic distributions of kick velocities and initial binary properties to determine the distributions of orbital parameters for HMXB progeneitors. In Section 3.3 we perform a similar analysis for LMXB progeneitors and in Section 3.4 we examine the HMXB progeneitor distributions for a restricted distribution of kick directions.

## 3.1 The dependence on $v_{\text{kick}}$

To explore the dependence of the post-supernova orbital parameters on the kick velocity, we consider two cases with fixed absolute kick velocities of $450\,\text{km}\,\text{s}^{-1}$ and $200\,\text{km}\,\text{s}^{-1}$. The distribution of kick directions is assumed to be isotropic in the frame of the neutron star. As in Section 2.3, we start with a zero eccentricity system with a $15\,M_\odot$ companion star and a $5\,M_\odot$ star that explodes in an asymmetric supernova leaving behind a $1.4\,M_\odot$ neutron star. We take the pre-kick orbital period distribution to be constant in the range between the minimum initial period at which star 2 would fill its Roche lobe and the maximum period for which the system remains bound (as determined by eq. 2.5). This is a serious and almost certainly unrealistic approximation, and we discuss and examine more realistic initial period distributions in the next subsection. Changes in the initial period distribution quantitatively affect the distributions of post-supernova orbital parameters, but not their qualitative features and the outer "envelope" of allowed quantities.

The distributions of orbital parameters as a function of post-supernova period are shown in Figs 5 and 6, respectively. In Fig. 7 we give the distributions of post-supernova orbital periods for all the simulations presented in this section. In Figs 5 and 6 we have added the additional constraint that, after the supernova has taken place, the radius of the companion star should not exceed its effective Roche lobe radius at periastron. Systems which violate this constraint suffer immediate dynamical mass transfer and lead to the possible merger of the two stars into a TŻO (Thorne & Żytkow 1975, 1977; Leonard, Hills & Dewey 1994). They are therefore removed from consideration. Such systems fall in the upper left hand corners of the two diagrams. In Fig. 5, 15 % of kicked systems remain bound and 42 % of these suffer strong dynamical mass transfer. In Fig. 6, 17 % of kicked systems remain bound and 15 % of these suffer strong dynamical mass transfer. There will be a few additional cases of strong dynamical interaction when a neutron star launched in what would be an unbound orbit collides with the companion star. Comparison of Fig. 5a and Fig. 6a shows that the relation between eccentricity and final orbital period becomes tighter as the kick velocity increases, since the possible range of initial periods and kick directions which leads to a bound system is more restricted. This result has potentially important consequences as we discuss below. The sharp rises in the distributions at around 18 days in Fig. 5a and 110 days in Fig. 6a occur around the maximum initial period at which circular post-supernova systems are possible (as given by eq. 2.9).

To compare the theoretical distributions with the properties of some observed HMXBs, we plot in these figures those X-ray binaries that have measured nonegligible eccentricities (the X-ray binaries $0114+650$, $0115+634$, $0236+610$, $0331+530$, $0535-668$, $0535+262$, $1145-619$, $1223-624$, $1907+097$, $2030+375$), in the hope that their orbital parameters have not evolved significantly since the supernova. This may not be the case for the LMC transient $0535-668$ which interestingly lies near our Roche lobe overflow curve and may therefore have evolved somewhat. We also plot the binary radio pulsars $0045-7319$ (Kaspi et al. 1994) and $1259-63$ (Johnston et al. 1994). The periods and eccentricities of these objects are listed in Table 1. If there has been no post-supernova orbital evolution, kicks of order $200\,\mathrm{km\,s^{-1}}$ can easily explain all measured systems while kicks of order $450\,\mathrm{km\,s^{-1}}$ predict higher minimum eccentricities than are seen in some of the long-period systems. However, our choices of kick and initial period distributions are hardly realistic and we will return to this issue after we have modified them appropriately. Van den Heuvel & Rappaport (1987) argue that kicks over and above those just due to anisotropic mass loss with respect to the center of mass (the so-called 'Blaauw' kicks; Blaauw 1961; Boersma 1961) are *required* to explain the observed eccentricities of Be X-ray binaries. This is consistent with our calculations. The period – eccentricity distributions for HMXBs with natal kick distributions constrained to lie entirely below $80\,\mathrm{km\,s^{-1}}$ (not shown) are completely inconsistent with the parameters of $0045-7319$, $0236+610$ and $0535-668$. These systems lie above the $80\,\mathrm{km\,s^{-1}}$ kick maximum eccentricity curve (for low kick velocities, the maximum eccentricity curve falls below the strong Roche-lobe overflow curve). Any post-supernova orbital circularization would only exacerbate this discrepancy. The lowest kick velocity that is marginally consistent with observed systems is $\sim 140\,\mathrm{km\,s^{-1}}$.

The distributions of the post-supernova system CM velocity relative to the initial CM velocity as a function of final orbital period are shown in Fig. 5b and Fig. 6b. As expected from eqs (2.7), (2.10) and (2.11), the final system velocities increase with the assumed kick velocity. For long final periods, the lowest possible

– 9 –

system velocity approaches the value given by the right-hand side of equation (2.11) (consider equations 2.7 and 2.10 in conjunction to understand this). Systems with short final periods have high CM velocities since short-period systems as well as systems with high CM velocities tend to be those in which the exploding star has received a kick more or less opposite to the direction of its motion (compare eqs 2.4 and 2.10).

In Fig. 5c and Fig. 6c, we plot the distributions of $\nu$ (eq. 2.16), the angle between the pre-supernova and post-supernova orbital-angular-momentum vectors. Since we take the spin vectors of the two stars to be aligned with the pre-supernova orbital-angular-momentum vector (see Sec. 2.1), $\nu$ measures the immediate post-supernova angle of the spin vectors of both stars with respect to the orbital-angular-momentum vector. It is clear that the fraction of systems where both stars have retrograde spins (i.e. where $\nu > 90°$) increases with increasing kick velocity. The sharp drops below final orbital periods of $\sim 4.2$ and $\sim 25$ days in Fig. 5c and Fig. 6c, respectively, occur since, in this parameter regime, post-supernova systems will experience immediate dynamical mass transfer and are therefore removed from the sample.

### 3.2 Post-supernova orbital parameters of HMXB progeneitors

Having illustrated the basic effects of kicks, we now refine our choices of kick and initial period distribution. We adopt the three-dimensional pulsar birth velocity distribution presented in fig. 2b of Lyne & Lorimer (1994) as our kick velocity distribution (this distribution has a mean value of $450 \, \mathrm{km \, s^{-1}}$). There is bias in doing this as some of the single pulsars measured by Lyne & Lorimer (1994) will have originated in binary star systems and high velocity kicks lead to more efficient ejection from binary systems (another effect which partially compensates this is that pulsars born in binary systems must climb out of the potential wells of their companions). We take the kick magnitude to be uncorrelated with binary properties, and this assumption needs further examination. Recent supernova computations which include convection show that rotation (which is coupled to binary properties) has an influence on the shape of the convective patterns that develop above the protoneutron star and may lead to its kick (section 5.5 of Herant et al. 1994). Further supernova computations which specifically examine this point are needed. In addition the kick magnitude may depend on the spatial opacity distribution and hence the metallicity of the progeneitor.

To compute the pre-supernova orbital period distribution, we consider either stable or dynamical mass transfer following the prescription given in section 3.1.1. of Podsiadlowski, Joss & Hsu (1992) (PJH). We take the pre-mass-transfer orbital-period distribution to be constant in $\log P$ between the periods obtained by demanding that there be no mass transfer on the main sequence as well as by demanding that mass transfer occur during later stellar evolution (see the discussion and references in PJH). In the case of stable mass transfer, we assume that half of the mass that is lost from the mass donor is lost from the system (e.g. De Greve 1992) carrying with it a specific angular momentum of order the systemic specific orbital angular momentum (i.e. we take $\alpha = 1$ and $\beta = 0.5$ in the formalism of PJH). In the case of dynamical mass transfer leading to a common-envelope (CE) phase, we assume that the CE-ejection process is very efficient and we set the CE efficiency parameter $\alpha_{\rm CE} = 1$ (e.g. Han, Podsiadlowski & Eggleton 1994). In our present exploratory investigation, we consider only one set of initial masses, $12\,M_\odot$ for the companion star and $15\,M_\odot$ for the star that will eventually be consumed in the supernova. Since these are very typical masses for a HMXB progenitor, we do not expect that our results would change significantly if we included a distribution of primary and secondary masses. With the PJH formalism, the mass of the companion star after the mass-transfer phase is either increased to $17\,M_\odot$ (in the case of stable mass transfer) or remains constant at $12\,M_\odot$ (in the case of dynamical mass transfer). The mass-losing star is transformed into a helium star of $5\,M_\odot$; since such a helium star does not become a helium giant subsequently, we do not need to consider a second phase of mass transfer (Habets 1986).

Fig. 8 shows the distributions of orbital properties as a function of final period for our refined kick and initial orbital period distributions. In this case, 27 % of systems remain bound and 26 % of these suffer strong dynamical mass transfer and are likely to become TŻOs. The 73 % of systems that become unbound will produce high velocity OB stars and this may then explain why most runaway OB stars are, apparently, single. Note that some of the fine structure in the distributions in Fig. 8 is due to the bimodality of the assumed companion mass in our mass transfer treatment and is therefore not necessarily real.

In Fig. 8a we see that, while the outer envelope of allowed eccentricities is fairly loose, the boundary within which 60 % of all systems lie is reasonably tight. Six of the twelve systems from Table 1 lie within the 60 % boundaries. For this small number of systems, this implies that the observed and theoretical distributions are roughly consistent with each other; but this agreement may be rather fortuitous, since our chosen systems cannot be considered an unbiased, representative sample. In addition, some of the orbital parameters of the plotted systems may have evolved since the supernova. Nevertheless, the comparison illustrates the potential of using the period – eccentricity distribution of eccentric neutron-star binaries to statistically deduce information about the distribution of supernova kicks and the pre-supernova orbital period distribution of HMXB progenitors after mass transfer (though this distribution lacks the some of the beauty of the analogous Phinney 1992 distribution). A proper analysis would have to be more sophisticated than the exploratory one presented here and should, for example, include estimates for the masses of individual systems (as estimated, e.g., from their spectral types) and consider a second phase of mass transfer, if appropriate. We note that there are $\approx 25$ systems with currently unmeasured periods, eccentricities, or both that could in principle be added to a diagram like Fig. 8a and that these represent only the tip of the $\sim 5000$ Galactic Be X-ray binary iceberg (van den Heuvel & Rappaport 1987). Systematic monitoring of Be X-ray binaries will add to the number of systems with measured orbital parameters (e.g. Coe et al. 1993; Roche et al. 1994), and X-ray sky monitors with large fields of view such as those on the BATSE, XTE, SAX and INTEGRAL satellites will find more systems. Deep X-ray imaging near supernova remnants and stellar nurseries will also find additional systems (e.g. Hughes & Smith 1994). Furthermore, if CM velocities can be determined as well (at least in a statistical sense), these can add additional constraints as per Fig. 8b.

In Fig. 8b we present the CM velocity distribution of post supernova HMXB progeneitors. As before, the CM velocity distribution rises towards low final orbital periods. It is worth noting that a reasonable fraction of our distribution lies below the $\sim$20–60 km s$^{-1}$ escape velocity from the core of a globular cluster (e.g. Pryor & Meylan 1993). Fig. 8c shows the spin tilt angle distribution. It is clear that most systems will become significantly misaligned by the supernova kick and that, in a significant fraction of systems (up to 20 %), both stars have retrograde spins after the supernova. We will discuss the possible effects of misalignment and retrograde spins in Section 5. Fig. 8d shows the distribution of the ratio of the companion star's spin angular momentum to the total system orbital angular momentum. To obtain a rough estimate of this quantity, we set the moment of inertia of star 2 equal to $\gamma M_2 R_2^2$ and use $\gamma \approx 0.1$ (e.g. Motz 1952). We assume that star 2 has been spun up during the first mass-transfer phase and is rapidly rotating at half its break-up rotation speed. For other choices of rotation speed and $\gamma$, the quoted ratios can be simply rescaled. As Fig. 8d shows, the companion star's spin angular momentum is a sizable fraction $(0.1-0.4)$ of the total orbital angular momentum. Thus, spin – orbit coupling effects and the Darwin instability (e.g. Pringle 1974 and references therein) may be important, especially in short-period systems (see Section 5).

The percentage of retrograde-spin systems is shown in Fig. 9a as a function of the final orbital period. We plot the percentages associated with Figs 5, 6 and 8 as well as their analog for a kick speed of 700 km s$^{-1}$. Fig. 9a illustrates how the fraction of systems with retrograde spins increases with increasing kick velocity. Note that, for $v_{\rm kick} = 700$ km s$^{-1}$, the fraction decreases with increasing orbital period, since all the short-period systems on bound prograde orbits experience immediate dynamical mass transfer and are removed. The fraction of systems with retrograde spins is greatly reduced in the most realistic case, using the Lyne-Lorimer distribution, for two reasons. The first is that the Lyne-Lorimer distribution contains many systems with relatively low velocities ($\lesssim 100$ km s$^{-1}$). Since these are more likely to remain bound after the supernova, they tend to dominate in the post-supernova sample. The second reason is that the more realistic pre-supernova period distribution favours systems with shorter periods which are more tightly bound and are less affected by the supernova kick. In Fig. 9b we show the fraction of systems with initially retrograde spins for the Lyne-Lorimer kick-velocity distribution as a function of systemic velocity for HMXBs and LMXBs (Section 3.3). It may be possible to use this strong correlation in some cases to infer from the systemic velocity, which in principle is an observable quantity, something that would not normally be observable (i.e. the orientation of the spin). For example, X Per which appears to be at a large height above the Galactic disk ($\sim 400$ pc for an assumed distance of 1300 pc; Fabregat et al. 1992) must have received a very large kick velocity as a result of the supernova, if it was born close to the Galactic disk. With a companion mass of $22\,M_\odot$ (Fabregat et al. 1992), it becomes very probable that the immediate post-supernova spins of the stars in X Per were retrograde with respect to the orbital motion. A similar situation may well apply for the probable LMXB Cir X-1.

### 3.3 Post-supernova orbital parameters of LMXB progeneitors

In Fig. 10 we show the orbital parameter distribution of LMXB progeneitors, assuming that these form similarly to cataclysmic variables, i.e. originate from binaries with a massive primary and a low-mass companion which experience a CE phase and a supernova explosion (for details see, e.g., Romani 1992). In this case, we take a typical pre-mass-transfer system consisting of a $13\,M_\odot$ primary and a $1\,M_\odot$ companion and assume that mass transfer is always dynamical. Thus, after mass transfer, the system consists of a $3.5\,M_\odot$ helium star with a $1\,M_\odot$ companion. As in Section 3.2, we adopt the Lyne-Lorimer distribution for the supernova kick distribution. With these assumptions, 19 % of systems remain bound after the supernova, of which 56 % experience immediate Roche-lobe overflow and, in most cases, lead to the destruction of the companion star (see the discussion in Section 5).

Fig. 10a shows that, for LMXB progeneitors, the immediate post-supernova period – eccentricity distribution is much tighter than the analogous HMXB distribution. The reason is that the constraints on the possible kick directions for which the binary systems remain bound are more restricted. Indeed, as is well known, without any kick, all systems with low-mass companions would become unbound (see Fig. 3a and the discussion in Bhattacharya & van den Heuvel 1991). For comparison, we plot in this diagram Cir X-1 ($P_{\rm orb} = 16.6$ days and $e > 0.7$), which probably is a LMXB, and the binary radio pulsar 2303+46 ($P_{\rm orb} = 12.34$ days and $e = 0.6584$) which has a $\sim 1.5\,M_\odot$ (compact) companion (Lyne & Bailes 1990) and whose properties before the last supernova would have been similar to those of a typical LMXB progenitor.

As is clear from Fig. 10a, all post-supernova systems with periods longer than a few days must initially be on eccentric orbits. Since it is unlikely that these systems will have circularized before the onset of mass transfer (as Cir X-1 proves), this suggests that there may a class of LMXBs analogous to Be X-ray binaries which experience X-ray outbursts when the companion fills (or possibly even overfills) its Roche lobe near periastron. Cir X-1 may be an extreme representative of this class. Since the transient X-ray outbursts would be strictly periodic, it should be possible to recognize these transients as a separate class with the present and future generation of all-sky X-ray monitors.

Fig. 10b shows that the system velocities of LMXBs are much larger than those of HMXBs (mainly because of the lower total mass of LMXBs). The average LMXB velocity is $180\,{\rm km\,s^{-1}}$ with a standard deviation of $80\,{\rm km\,s^{-1}}$. Since this velocity is comparable to the Galactic rotation velocity ($220\,{\rm km\,s^{-1}}$), this immediately suggests that the space distribution of LMXBs can evolve substantially from the space distribution of their progenitors (see Section 4.2). Our distribution is marginally consistent with PSR J1713+0747, which has a current orbital period of 67.8251 days, a companion mass in the range 0.27–0.4 $M_\odot$ and a probable initial velocity of less than 80 km s$^{-1}$ (Camilo et al. 1994).

On the other hand, the orientation of the orbital plane of a post kick LMXB progeneitor is not dramatically changed by the supernova kick, and the post-supernova tilt of the spin axis with respect to the orbital axis is relatively small (see Fig. 10c). Since in addition the ratio of the spin to orbital angular momentum is very small in LMXBs (Fig. 10d), spin – orbit coupling effects are unlikely to be important for LMXBs.

### 3.4 Angularly restricted kick distributions

So far, we have always assumed that the distribution of kick directions is isotropic in the neutron star's frame. However, in some theoretical models for supernova kicks, one might expect the kick to be preferentially along the initial spin axis. To examine the consequences of this possibility, we also calculated the post-supernova orbital distribution for a case in which we restricted the range of kick directions to a cone along the initial spin axis of the exploding star (assumed to be aligned with the orbital axis) with an opening angle of 20°. All other parameters are the same as in Section 3.2. The results of these simulations are presented in Fig. 11.

As the comparison of Figs 11a, 11b and 11d with the corresponding Figs 8a, 8b and 8d shows, the eccentricity – orbital period, the system velocity – orbital period and the spin/orbital angular momentum ratio – orbital period distributions are qualitatively very similar in the two cases. However, in the case of restricted kick directions, the upper eccentricity boundary is determined by the restricted range of kick directions and not by the condition that the post-supernova system experiences immediate Roche-lobe overflow. Consequently, the upper eccentricity boundary lies somewhat lower in Fig. 11a than in Fig. 8a and falls below the point for 0535-668. This may provide some weak evidence against a very restricted range of kick directions. Another consequence is that, in the case of very restricted kick directions, there are no systems that experience immediate dynamical mass transfer and lead to the immediate formation of a TŻO. The number of systems that remain bound after the supernova is also somewhat reduced from 27 % to 22 %. The distribution of tilt angles (Fig. 11c) is dramatically different. With the restricted kick directions, the possible range of spin tilt angles becomes very restricted, now strongly favoring low angles and systems with prograde spins.

### 4 SKY DISTRIBUTIONS

In Section 3 we have shown that X-ray binaries receive a significant kick because of an asymmetric supernova explosion. This will affect their space distribution.

– 14 –

Indeed, the observed space distribution may allow inferences about the required kick velocity.

To compare our results with observed sky distributions, we have performed a series of Monte-Carlo simulations in which we calculate the trajectories of $\sim 400\,000$ binary systems which receive a kick when the first star explodes and determine the resulting sky distributions. Following a similar analysis for gamma-ray bursters by Paczyński (1990), we assume that the initial radial distribution, $P_r(r)$, of binaries follows the disk distribution of stars, i.e.

$$P_r(r)\,\mathrm{d}r \propto r\,\exp(-r/r_d)\,\mathrm{d}r, \qquad (4.1)$$

where $r$ is the distance from the Galactic center and $r_d$ the disk scale length, taken to be $4.5\,\mathrm{kpc}$ (van der Kruit 1987). In most simulations, we cut off the distribution at $5\,R_0$, where $R_0$ is the distance from the Sun to the Galactic center, assumed to be $8\,\mathrm{kpc}$. For the initial distribution perpendicular to the disk, $P_z(z)$, we adopt an exponential distribution

$$P_z(z)\,\mathrm{d}z \propto \exp(-|z|/z_d)\,\mathrm{d}z, \qquad (4.2)$$

where $z$ is the height above the midplane of the Galactic disk and $z_d$ the disk scale height for massive stars, taken to be $75\,\mathrm{pc}$ (van der Kruit 1987). As a Galactic potential we use the potential of Miyamoto & Nagai (1975) in the form given by Paczyński (1990). This potential assumes a circular rotation velocity of $220\,\mathrm{km\,s^{-1}}$ at the solar circle.

### 4.1 The sky distribution of HMXBs

The observed sky distribution of HMXBs is strongly concentrated towards the Galactic disk. To quantify this, we define two moments of the distribution (similar to the common practice in the study of gamma-ray bursts), a dipole moment $\langle \cos\theta \rangle$, where $\theta$ is the angle between a system and the Galactic center, and a quadrupole moment $\langle \sin^2 b \rangle$, where $b$ is the Galactic latitude of a system. We calculated these moments using the compilation of HMXBs by van Paradijs (1994), where we discarded some systems whose massive nature is uncertain or unlikely (e.g. in the case of some Be systems of late-B spectral type). The dipole and quadrupole moments of the selected 54 systems are 0.24 and 0.0042, respectively. We note that the relatively large value of the quadrupole moment is mainly caused by two systems at relatively large Galactic latitudes, X Per and 1936+541. Excluding both of these systems would reduce the quadrupole moment drastically to 0.0013. On the other hand, X Per is certainly a massive system and should be included, while the nature of 1936+541 is less certain. Excluding only 1936+541 would give a quadrupole moment of 0.0029.

Another problem with this observed distribution is that it is obviously very incomplete, as can be seen from the strong relative deficiency of systems in the direction of the Galactic center. We find that we can reasonably well account for these selection effects if we assume that the observed systems have distances larger than 4.8 kpc from the Galactic center and are within 8 kpc of the Sun (the latter constraint is not very strong). With these assumptions, we obtain a longitudinal distribution of HMXBs which is very similar to the observed one. Having thus fixed the longitudinal distribution, we can examine the latitudinal distribution of HMXBs, which will be mainly a function of the distribution of kick velocities and the characteristic age of HMXBs.

We use the distribution of post-supernova system velocities calculated in Section 3.2 which assumed a Lyne-Lorimer distribution for the neutron-star kick velocities. This distribution has an average velocity of $51\,\mathrm{km\,s^{-1}}$ with a standard deviation of $18\,\mathrm{km\,s^{-1}}$. In Table 2a we present the predicted dipole and quadrupole moments and the average height, $\langle |z| \rangle$, of HMXBs above (or below) the disk for two simulations in which we assumed that the maximum age of HMXBs since the supernova, $t_{\max}$, is $5\times 10^6$ and $10^7$ yr, respectively. We assume that all ages up to $t_{\max}$ are equally probable and did not include a possible *systematic* delay between the initial supernova and the beginning of the X-ray phase. This assumption is probably justified for the majority of typical Be X-ray binaries (which dominate the sample), but is invalid for HMXBs with Roche-lobe filling or evolved mass donors. The quoted uncertainties in Table 2a for the dipole and quadrupole moments are the standard deviations from the expectation values for a realization of the underlying distribution with 54 systems (the same as our observational sample). Considering the small observational sample, the uncertainties due to the observational selection effects and the simplicity of our model, the only thing we can conclude from Table 2a is that the simulated distributions are roughly consistent with the observed sample. To illustrate this graphically, we present in Fig. 12a a realization of 54 systems for the $5 \times 10^6$ yr simulation.

### 4.2 The sky distribution of LMXBs

To examine the sky distribution of LMXBs, we use the sample of flux-limited LMXBs compiled by Naylor & Podsiadlowski (1993). The advantage of this sample, containing 36 systems, is that it was designed to be more or less complete and that we do not have to worry about selection effects. The dipole and quadrupole moments of this sample are 0.76 and 0.024, respectively. In our LMXB simulations we use the distribution of system velocities calculated in Section 3.3, which again assumed a Lyne-Lorimer neutron-star kick velocity distribution. This LMXB distribution has an average system velocity of $180\,\mathrm{km\,s^{-1}}$ with a standard deviation of $80\,\mathrm{km\,s^{-1}}$. In Table 2b we present the results of six simulations, the first five of which illustrate the evolution of the dipole and quadrupole moments and the average height above the disk as a function of the maximum time since the supernova, $t_{\max}$. Note that we again did not include a possible *systematic* time delay between the supernova and the beginning of the LMXB phase. If the characteristic delay is a significant fraction of $t_{\max}$, the dipole moment would be reduced and the quadrupole moment and average disk height increased relative to the values given in

Table 2b. Fortunately, even if this effect were important, it would not significantly affect our main conclusions below.

Table 2b illustrates several important points. As $t_{\max}$ increases, both the quadrupole moment and the average $z$-height increase as expected. After $\sim 10^8$ yr, the $z$ distribution has more or less reached its maximum extent and only changes marginally afterwards. Slightly more surprising is that the dipole moment decreases significantly with time. The reason is that the average system kick velocity ($\sim 180\,\mathrm{km\,s^{-1}}$) is comparable to the Galactic rotation velocity ($220\,\mathrm{km\,s^{-1}}$) and that the distribution of LMXBs therefore spreads out significantly away from the Galactic center. The theoretical moments are roughly consistent with the observed ones for all $t_{\max} \gtrsim 2 \times 10^7$ yr (the quoted uncertainties in Table 2a give the deviations from the expectation values for realizations of 36 systems). The dipole moment for $t_{\max} \gtrsim 10^8$ yr is marginally inconsistent with the observed moment (at the 1.4-$\sigma$ level). This may provide a weak suggestion that LMXBs are relatively young with ages less than $\sim 10^8$ yr. An alternative explanation is that the initial radial distribution (eq. 4.1) of LMXBs has a cut-off at some radial distance $r_{\max}$. This is illustrated by the sixth simulation in Table 2b where we assumed that the initial radial distribution is cut off at the solar circle. The resulting moments are consistent with the observed ones even for $t_{\max} \gtrsim 10^8$ yr. A cut-off at the solar circle is close to the observed cut-off of molecular clouds (Gilmore 1994). Since molecular clouds mark the regions of current massive star formation, this may also suggest that LMXBs are descendants of a relatively young population of stars.

The main and most important conclusion of our LMXB simulations is that the observed Galactic LMXB distribution is consistent with a normal Galactic disk population which has been widened because LMXBs receive a significant kick at birth and that LMXBs do not require a special population of progenitors (see also Bailes [1989] and Naylor & Podsiadlowski [1993] for similar conclusions). In Fig. 12b we present a typical realization of 36 LMXBs for the simulation with $t_{\max} = 2 \times 10^7$ yr, which is remarkably similar to the observational sample (compare figure 1 in Naylor & Podsiadlowski [1993]). The observed distribution does not provide a strong constraint on the maximum ages of LMXBs since the supernova, with all ages between $\sim 2 \times 10^7$ and $10^{10}$ yr being possible. There may, perhaps, be a weak hint that the characteristic age of LMXBs may be relatively young ($t_{\max} \lesssim 10^8$ yr).

## 5 DISCUSSION

One of the main results of this study is that the observed orbital parameters and the sky distributions of neutron-star binaries can be well reproduced with the new increased supernova kick velocities. In addition, we have shown that there are some fairly good correlations between different orbital parameters for HMXB and LMXB progenitors (for example, orbital period and eccentricity, orbital period and system velocity), which can be used to provide information about the distribution of supernova kick velocities, the distribution of kick directions and the pre-supernova orbital period distribution. We expect that these correlations may be fruitfully exploited in the future as the sample of eccentric neutron-star binaries (e.g. Be X-ray binaries) with well-determined orbital parameters increases and becomes more representative.

Because of the large kick velocity, a significant fraction of bound post-supernova systems ($\sim 1/4$ of systems with massive companions, $\sim 1/2$ of systems with low-mass companions) will experience immediate dynamical mass transfer. If the companion is a massive star, it is likely to lead to the immediate spiral-in of the neutron star and the possible formation of a TŻO. If the companion is a low-mass star, the consequences of immediate Roche-lobe overflow may depend on the direction of the kick. If the neutron star collides directly with its companion star, the formation of a low-mass TŻO is possible. If there is no direct collision, but the companion star overfills its Roche lobe by a large amount, the companion star is likely to be dynamically disrupted, probably leading to the formation of a massive disk around the neutron star. The X-ray system 1E2259+586 in the supernova remnant G109.1-1.0 may be such a system. Despite the fact that it has the appearance of a LMXB and seems to accrete from an accretion disk, there is no sign of a binary companion. This is particularly puzzling since it is a X-ray pulsar, which should make the detection of a binary companion relatively easy. Neutron stars surrounded by massive disks are good candidates for the progenitors of neutron stars with planetary systems (Stevens, Rees & Podsiadlowski 1992). If the companion star overfills its Roche lobe only during a short phase near periastron, it may not be dynamically disrupted. In this case, it should exhibit dramatic periodic outbursts near periastron. Cir X-1 may provide a good example for such a system. This interpretation becomes particularly probable, if it is indeed associated with the supernova remnant $G321.9 - 0.3$ (Haynes et al. 1978; Haynes et al. 1986; Stewart et al. 1993), which would prove its relative youth. It is worth noting that the ephemeris given in Glass (1994) suggests a orbital decay timescale of $\sim 5000$ years if it is interpreted in the simplest manner possible.

Another important result is that our simulations predict that, in most massive neutron-star binaries, the immediate post-supernova spin axes have large inclinations with respect to their orbital axes (the average tilt is $\sim 40°$ and the median is $\sim 20°$) and that a significant fraction of systems (up to $\sim 20\,\%$) have retrograde spins. This may have important consequences for the appearance and long-term behavior of HMXBs (see Hills [1983] for a detailed previous discussion). For example, it suggests that the disk-like winds from the Be stars in Be X-ray binaries are typically not confined to the orbital planes. This will significantly affect the geometry of the accretion process onto the neutron star and the accretion rate. The latter depends to a relatively high (third) power on the relative velocity of the neutron star with respect to the Be-star wind (see also Cook & Warwick 1987). If the Be star precesses with respect to the orbital axis, this could introduce a long-term, possibly periodic, cycle in some Be X-ray binaries. This may be observed, for example, in 0236+610, which seems to exhibit a long-term, periodic ($\sim 1600\,\mathrm{days}$) cycle (Ribas 1993). It may also account for the long-term decline in Cir X-1 (Whitlock & Tyler 1994).

A misaligned companion star may have even more important consequences in systems in which accretion onto the neutron star occurs through an accretion disk. Precession of a misaligned companion star has been suggested as a plausible mechanism to explain the precessing, tilted and twisted accretion disk which may be seen in some X-ray binaries (e.g. Roberts 1974; Petterson 1977). Of course, as emphasized by Papaloizou & Pringle (1982), stars are gaseous bodies and do not precess like solid bodies. If the spectrum of excited rotational oscillation modes is continuous and broad, the precessional motion of a star is likely to be damped out completely. However, if a few modes dominate, the star's motion may resemble precession, although it is unlikely to be as regular as the precession of a solid body. These effects will be even more dramatic if the spin is retrograde with respect to the orbital motion. In this case, simple spin-orbit alignment may be impossible (indeed tidal effects may favor counter-alignment) and the spin-orbit coupling will cause the long-term decay of both the spin period and the orbital period. This may be an additional mechanism to drive orbital decay in some HMXBs (e.g., the Cen X-3/Krzemiński's star system; Pringle 1974; Kelley et al. 1983). Work is needed to extend the results of Papaloizou & Pringle (1982) to larger precessional amplitudes so that these problems may be addressed. The fact that the spin angular momentum in the massive component of most HMXBs is a significant fraction of the total orbital angular momentum implies that the orbital-angular-momentum vector will precess and that it is not correct to consider the orbital plane of the system as fixed in time. This may produce subtle but observable effects such as variations in eclipse durations in eclipsing systems and variations in X-ray pulse profiles.

In a significant fraction of HMXBs, the neutron stars will initially possess retrograde spins. Accretion onto such stars will cause them to spin down. This may in part account for the low spin rates observed in many X-ray pulsars (Hills 1983), although it is unlikely to be the whole explanation, since very little accretion after the spin-down phase would be sufficient to spin up the neutron stars in a prograde sense. Nevertheless, the accretion process itself should be affected by a retrograde neutron-star spin, for example, by increasing the accretion efficiency (see Hills 1983).

In the massive radio pulsar binary PSR J0045-7319 (Kaspi et al. 1994), it may be possible to measure the sense of the spin of the massive companion star by its effect on the apsidal advance (e.g. Smarr & Blandford 1976). A retrograde spin would decrease the apsidal advance, which could explain the relatively low upper limit measured by Kaspi et al. (1994).

Podsiadlowski et al. (1993) have argued that the progenitor of supernova 1993J (SN 1993J) was a member of an interacting binary with an immediate pre-supernova period of $\sim 3000$ days and suggested that this system may become a X-ray binary in the future. However, for such a wide system to remain bound after the supernova, equation (2.5) implies that the supernova kick velocity must be strictly less than $\sim 90\,\mathrm{km\,s^{-1}}$ (for typical parameters appropriate for SN 1993J). In view of the increased pulsar birth velocities, it now appears more likely that the system will have become unbound. On the other hand, if the system has remained bound, its post-supernova orbital parameters would provide strong constraints on the magnitude and direction of the supernova kick.

As our discussion shows, the new increased supernova kick velocities have significant consequences for the evolution and appearance of X-ray binaries and related systems. While a number of these have been investigated in the past, other have not and certainly deserve further exploration in the future.

**ACKNOWLEDGEMENTS**

We thank S.T. Fabian and P.W.D. Johnstone, whose high velocity natal kicks inspired this work. We thank A.G. Lyne and D.R. Lorimer for providing us with their distribution of velocities of young pulsars, S. Sigurdsson and R. Wijers for helpful discussions, an anonymous referee for suggestions, and R. Johnstone and D. White for installing the Pascal compiler on our computer system. WNB gratefully acknowledges support from the National Science Foundation of the United States of America and the British Overseas Research Studentship Programme. This research has made use of data obtained from the SIMBAD database, operated at CDS, Strasbourg, France.

**REFERENCES**

Amnuel P. R., Guseinov O. H., 1976, A&A, 46, 163

Anderson B., Lyne A. G., 1983, Nat, 303, 597

Bailes M., 1989, ApJ, 342, 917

Bailes M., Manchester R. N., Kesteven M. J., Norris R. P., Reynolds J. E., 1989, ApJ, 343, L53

Bhattacharya D., van den Heuvel E. P. J., 1991, Phys. Rep., 203, 1

Blaauw A., 1961, Bull. Astr. Inst. Netherlands, 15, 265

Boersma J., 1961, Bull. Astr. Inst. Netherlands, 15, 291

Burrows A., Fryxell B. A., 1992, Science, 258, 430

Camilo F., Foster R.S., Wolszczan A., 1994, ApJ, in press

Caraveo P. A., 1993, ApJ, 415, L111

Coe M. J., Everall C., Fabregat J., Gorrod M.J., Norton A. J., Reglero V., Roche P., Unger S. J, 1993, A&AS, 97, 245

Colgate S., 1970, Nat, 225, 247

Cook M. C., Warwick R. S., 1987, MNRAS, 225, 369

Crampton D., Hutchings J. B., Cowley A. P., 1985, ApJ, 299, 839

De Cuyper J.P., 1984, Ph.D. thesis, Vrije Universiteit Brussels

De Cuyper J.P., De Greve J.-P., De Loore C., van den Heuvel E.P.J., 1976, A&A, 52, 315

De Greve J.-P., 1992, in Kondo Y., Sisteró R. F., Polidan R. S., eds, Evolutionary Processes in Interacting Binary Stars. Kluwer, Dordrecht, p. 41

Dewey R. J., Cordes J. M., 1987, ApJ, 321, 780

Duncan A. R., Stewart R. T., Haynes R. F., 1993, MNRAS, 265, 157

Duncan R. C., Thompson C., 1992, ApJ, 392, L9

Eggleton P. P., 1983, ApJ, 268, 368

Fabregat J. et al., 1992, A&A, 259, 522

Finger M. H., Cominsky L. R., Wilson R. B., Harmon B. A., Fishman G. J., 1994, in Holt S. S., Day C. S., eds, The Evolution of X-ray Binaries. AIP Press, New York, p. 459

Fomalont E. B., Goss W. M., Lyne A. G., Manchester R. N., Justtanont K., 1992, MNRAS, 258, 497

Flannery B.P., van den Heuvel E.P.J., 1975, A&A, 39, 61

Frail D. A., Goss W. M., Whiteoak J. B. Z., 1994, MNRAS, submitted

Fryxell B. A., Arnett W. D., 1981, ApJ, 243, 994

Gilmore G., 1994, private communication

Giovannelli F., Graziati L. S., 1992, Space Sci Rev, 59, 1

Glass I.S., 1994, MNRAS, 268, 742

Habets G. M. H. J., 1986, A&A, 167, 61

Habets G. M. H. J., 1987, A&A, 184, 209

Han Z., Podsiadlowski Ph., Eggleton P. P., 1994, MNRAS, submitted

Harrison E. R., Tademaru E., 1975, ApJ, 201, 447

Harrison P. A., Lyne A. G., 1993, MNRAS, 265, 778

Harrison P. A., Lyne A. G., Anderson B., 1993, MNRAS, 261, 113

Haynes R. F., Jauncey D. L., Murdin P. G., Goss W. M., Longmore A. J., Simons L. W. J., Milne D. K., Skellern D. J., 1978, MNRAS, 185, 661

Haynes R. F. et al., 1986, Nat, 324, 233

Herant M., Benz W., Colgate S. A., 1992, ApJ, 395, 642

Herant M., Benz W., Hix W.R., Fryer C.L., Colgate S. A., 1994, ApJ, 435, 339

Hills J. G., 1983, ApJ, 267, 322

Hughes J. P., Smith R. C., 1994, ApJ, in press

Hutchings J. B., Crampton D., 1981, PASP, 93, 486

Hutchings J. B., Crampton D., Cowley A. P., Olszewski E., Thompson I. B., Suntzeff N., 1985, PASP, 97, 418

Janka H.-T., Müller E., 1994, A&A, submitted

Johnston S., Manchester R.N., Lyne A.G., Nicastro L., Spyromilio J., 1994, MNRAS, 268, 430

Kaspi V. M., Johnston S., Bell J. F., Manchester R. N., Bailes M., Bessell M., Lyne A. G., D'Amico N., 1994, ApJ, 423, L43

Kelley R. L., Rappaport S., Brodheim M. J., Cominsky L., Stothers R., 1981, ApJ, 251, 630

Kelley R. L., Rappaport S., Clark G. W., Petro L. D., 1983, ApJ, 268, 790

Kulkarni S. R., Frail D. A., Kassim N. E., Murakami T., Vasisht G., 1994, Nat, 368, 129

Leonard P. J. T., Hills J. G., Dewey R. J., 1994, ApJ, 423, L19

Lyne A. G., Bailes M., 1990, MNRAS, 246, 15p

Lyne A. G., Lorimer D. R., 1994, Nat, 369, 127

Makishima K., Kawai N., Koyama K., Shibazaki N., Nagase F., Nakagawa M., 1984, PASJ, 36, 679

Miyamoto M., Nagai R., 1975, PASJ, 27, 533

Motz L., 1952, ApJ, 115, 562

Naylor T., Podsiadlowski Ph., 1993, MNRAS, 262, 929

Paczyński B., 1990, ApJ, 348, 485

Papaloizou J. C. B., Pringle J. E., 1982, MNRAS, 200, 49

Petterson J. A., 1977, ApJ, 216, 827

Phinney E.S., 1992, Philos. Trans. R. Soc. London Ser. A, 341, 39

Podsiadlowski Ph., 1989, Ph.D. thesis, Massachusetts Institute of Technology

Podsiadlowski Ph., Hsu J. J. L., Joss P. C., Ross R. R., 1993, Nat, 364, 509

Podsiadlowski Ph., Joss P. C., Hsu J. J. L., 1992, ApJ, 391, 246

Pringle J.E., 1974, MNRAS, 168, 13P

Pryor C., Meylan G., 1993, in Djorgovski S.G., Meylan G., eds, Structure and Dynamics of Globular Clusters. PASP Press, San Francisco, p. 357

Ribas J. M., 1993, Ph.D. thesis, Universitat de Barcelona

Ricketts M. J., Hall R., Page C. G., Pounds K. A., 1981, Space Sci Rev, 30, 399

Roberts W. J., 1974, ApJ, 187, 575

Roche P. et al., 1994, in Holt S. S., Day C. S., eds, The Evolution of X-ray Binaries.

    AIP Press, New York, p. 487

Romani R. W., 1992, ApJ, 399, 621

Rothschild R. E., Kulkarni S. R., Lingenfelter R. E., 1994, Nat, 368, 432

Sato N., Nagase F., Kawai N., Kelley R. L., Rappaport S., White N. E., 1986, ApJ, 304, 241

Shklovskii I. S., 1969, Sov. Ast., 13, 562

Skinner G. K., 1981, Space Sci Rev, 30, 441

Smarr L. L., Blandford R., 1976, ApJ, 207, 574

Stella L., White N. E., Davelaar J., Parmar A. N., Blissett R. J., van der Klis M., 1985, ApJ, 288, L45

Stevens I. R., Rees M. J., Podsiadlowski Ph., 1992, MNRAS, 254, 19p

Stewart R. T., Caswell J. L., Haynes R. F., Nelson G. J., 1993, MNRAS, 261, 593

Stollberg M. T., Paciesas W. S., Finger M. H., Fishman G. J., Wilson R. B., Harmon B. A., Wilson C. A., 1994, in Holt S. S., Day C. S., eds, The Evolution of X-ray Binaries. AIP Press, New York, p. 255

Sutantyo W., 1978, Astrophys. Sp. Sc., 54, 479

Sutantyo W., van der Linden T. J., van den Heuvel E. P. J., 1986, A&A, 169, 133

Taylor J. H., Cordes J. M., 1993, ApJ, 411, 674

Thorne K. S., Żytkow A. N., 1975, ApJ, 199, L19

Thorne K. S., Żytkow A. N., 1977, ApJ, 212, 832

van den Heuvel, E. P. J., 1994, in Vanbeveren D., van Rensbergen W., de Loore C., eds, Evolution of Massive Stars: A Confrontation between Theory and Observation. Kluwer, Dordrecht, p. 309

van den Heuvel, E. P. J., Rappaport S., 1987, in Slettebak A., Snow T.P., eds, Physics of Be Stars. Cambridge Univ Press, Cambridge, p. 291

van der Kruit P. C., 1987, in Gilmore G., Carswell B., eds, The Galaxy, Series C: Mathematical and Physical Sciences, Vol. 207. Reidel, Dordrecht, p. 27

van Paradijs J., 1994, in Lewin W. H. G., van Paradijs J., van den Heuvel E. P. J., eds, X-ray Binaries. Cambridge University Press, Cambridge, in press

Verbunt F., Wijers R. A. M. J., Burm H. M. G., 1990, A&A, 234, 195

Wheeler J. C., Lecar M., McKee C. F., 1975, ApJ, 200, 145

Whitlock L. & Tyler P., 1994, Legacy #4, The Journal of the High Energy Astrophysics Science Archive, NASA/GSFC

Wijers R. A. M. J., van Paradijs J., van den Heuvel E. P. J., 1992, A&A, 261, 145

Woosley S. E., 1987, in Helfand D. J., Huang J.-H., eds, The Origin and Evolution of Neutron Stars. Reidel, Dordrecht, p. 255

## Table Captions

**Table 1.** Orbital periods and eccentricities for the neutron-star binaries shown in Figs 5a, 6a, 8a and 11a. Another useful table of this type is given in Habets (1987).

**Table 2.** Dipole and quadrupole moments and average disk heights for simulated sky distributions of HMXBs (a) and LMXBs (b) with various maximum ages since the supernova (as indicated).

# Figure Captions

**Figure 1.** Sketch showing the binary system and supernova kick geometry. The angle $\theta$ specifies the angle between the direction of the kick-velocity vector and the initial orbital plane, and $\phi$ is the angle between the initial direction of motion of star 1 and the projection of the kick-velocity vector onto the orbital plane.

**Figure 2.** The probability for bound orbits as a function of $v_{\rm kick}/v_{\rm orb}$ for various values of $\tilde{m}$, as indicated.

**Figure 3.** Regimes for bound post-supernova systems and systems with prograde and retrograde spins in a $\tilde{v} - \tilde{m}$ plane (a) and $\log P_{\rm orb} - v_{\rm kick}$ plane (b), respectively. Fig. 3b assumes a total initial mass of $20\,M_\odot$ and a total post-supernova mass of $16.4\,M_\odot$. The degree of shading indicates the fraction of systems with retrograde spins (from left to right: all prograde, less than 50 % retrograde, more than 50 % retrograde, all retrograde).

**Figure 4.** Distributions of (a) final orbital period, (b) final eccentricity and (c) center-of mass (CM) velocity as functions of *initial* orbital period for a typical HMXB progeneitor in which the newborn neutron star received a kick of $450\,{\rm km\,s^{-1}}$ for an isotropic distribution of kick directions. Other system parameters are described in the text. The central solid curves show the medians of the distributions. Curves moving progressively outward from these medians include 20, 40, 60, 80 and 98 % of all systems (at a fixed initial period). The region which includes 60 % of systems has been shaded. The solid vertical lines on the left and the right mark the period at which star 2 fills its Roche lobe before the supernova and the maximum period which leads to bound post-supernova systems. The other vertical lines, from left to right, show the minimum period for which post-supernova systems with retrograde spins are possible, the period at which 50 % have retrograde spins and the period above which all systems have retrograde spins.

**Figure 5.** Distributions of (a) final eccentricity, (b) system CM velocity and (c) angle between the pre-supernova and post-supernova orbital-angular-momentum vectors as functions of the *final* orbital period for a typical HMXB progeneitor. The simulations assume an isotropic kick distribution with a fixed absolute kick velocity of $450\,{\rm km\,s^{-1}}$. Other system parameters are explained in the text. The distribution demarcation lines are as in Fig. 4 and the dashed vertical line shows the final orbital period at which half of the post-supernova systems have retrograde spins (note that this orbital period is longer than the orbital period implied by eq. 2.22, since the latter does exclude systems that experience immediate dynamical

mass transfer). The region which includes 60 % of systems has been shaded. The degree of shading indicates the relative fraction of systems as a function of final orbital period. In Fig. 5a we plot, for comparison, several eccentric X-ray binaries (filled circles and upward arrows — see Table 1) and two binary radio pulsars with massive companions (stars). The median final orbital period in the simulation is 28.64 days; the average eccentricity, CM velocity and tilt angle are 0.63, 79.8 km s$^{-1}$ and 119°, respectively.

**Figure 6.** Same as Fig. 5 but for an isotropic 200 km s$^{-1}$ kick distribution. The median final orbital period in the simulation is 144.1 days; the average eccentricity, CM velocity and tilt angle are 0.68, 40.0 km s$^{-1}$ and 97°, respectively.

**Figure 7.** Number of systems as a function of final orbital period for the simulations in Figs. 5 (dotted curve), 6 (dot-dashed curve), 8 (solid curve), 10 (dashed curve) and 11 (triple-dot-dashed curve). The total number of systems for each curve (i.e. the integral $\int N(P) \, d\log P$) has been normalized to 100. As these curves are for HMXB and LMXB progeneitors before the accretion associated with X-ray activity has begun, there may be further orbital period evolution before accretion starts (e.g. note the peak period of our LMXB distribution is larger than that typically observed).

**Figure 8.** Distributions of (a) final eccentricity, (b) system CM velocity, (c) angle between the pre-supernova and post-supernova angular-momentum vectors and (d) spin to orbital angular momentum ratio as functions of final orbital period for a typical HMXB. An isotropic kick distribution has been used where the kick speeds are drawn from the three-dimensional distribution of Lyne & Lorimer (1994). The pre-supernova period distribution and other parameters are described in the text. The distribution demarcation lines and the shading are as in Fig. 4. X-ray binary and binary radio pulsar systems are plotted as in Fig. 5a. The median final orbital period in the simulation is 25.6 days; the average eccentricity, CM velocity, tilt angle and spin to angular momentum ratio are 0.56, 51.2 km s$^{-1}$, 39° and 0.23, respectively.

**Figure 9.** (a) Percentage of systems with retrograde spins as a function of final orbital period for various kick velocity distributions (as indicated) and (b) as a function of post-supernova center-of-mass velocity for a Lyne-Lorimer velocity distribution for HMXBs and LMXBs.

**Figure 10.** Same as Fig. 8 for a typical LMXB. The median final orbital period in the simulation is 2.9 days; the average eccentricity, CM velocity, tilt angle and spin to angular momentum ratio are 0.58, 180.0 km s$^{-1}$, 27° and 0.02, respectively.

**Figure 11.** Same as Fig. 8, except that the allowed range of kick directions has been restricted to a cone along the spin axis with an opening of 20°. The median final orbital period in the simulation is 29.4 days; the average eccentricity, CM velocity, tilt angle and spin to angular momentum ratio are 0.53, $52\,\mathrm{km\,s^{-1}}$, 29° and 0.22, respectively.

**Figure 12.** Simulated sky distributions for a sample of 54 HMXBs (a) and 36 LMXBs (b). The initial system velocity distributions were taken from the simulations in Section 3.2 and 3.3 (see Figs 8c and 10c). The simulations assume that systems can have all ages with equal probability since the supernova up to a maximum age of $5\times10^6$ yr (a) and $2\times10^7$ yr (b), respectively. The dipole and quadrupole moments of the realizations are 0.79 and 0.025 (a) and 0.29 and 0.0025, respectively.